# Magnetic properties of a spin-1 Triangular Ising system

**Ersin Kantar**[1]**, Yusuf Kocakaplan**[2] **and Mehmet Ertaş**[1,*]

[1]*Department of Physics, Erciyes University, 38039 Kayseri, Turkey*
[2]*Graduate School of Natural and Applied Sciences, Erciyes University, 38039 Kayseri, Turkey*

**Abstract**

We studied some magnetic behaviors of Blume-Capel (BC) model in a site diluted triangular lattice by means of the effective-field theory (EFT) with correlations. The effects of the exchange interaction (J), crystal field (D), concentration (p) and temperature (T) on the magnetic properties of spin-1 BC model in a triangular lattice such as magnetization, susceptibility, phase diagram and hysteresis behaviors are investigated, in detail. The phase diagrams of the system are presented in two different planes. The tricritical point as well as tetracritical and critical end special points are found as depending on the physical parameters of the system. Moreover, when the hysteresis behaviors of the system are examined, the single and double hysteresis loop are observed for various values of the physical parameters. We show that the hysteresis loops have different coercive field points in which the susceptibility make peak at these points.

## 1. INTRODUCTION

Recently, the triangular lattice structure increasingly attractive more interest due to the fact that it has been prototypical modeling to the some real magnetic materials, such as $CsCoX_3$ (X = Cl or Br) and $Ca_3Co_2O_6$ spin-chain compounds, and for low-temperature and high-degenerate case show different behaviors [1-3]. In triangular lattice, local geometric constraints prevent simultaneous minimization of all the pairwise interactions. Thus, the Ising model with nearest-neighbor antiferromagnetic interactions on a triangular lattice is fully frustrated [4]. First-order [5] and second-order [6] phase transitions features of the triangular Ising model with nearest- and next-nearest-neighbor antiferromagnetic interactions have been studied by using a Wang-Landau entropic sampling scheme. By the utilizing EFT with correlations, Žukovič *et al.* [7] have investigated magnetization processes and phase transitions in a geometrically frustrated triangular lattice Ising antiferromagnet in the presence of a random site dilution and an external magnetic field. Melchert and Hartmann [8] have presented an algorithm for the computation of ground state spin configurations for the 2$d$ random bond Ising model on triangular lattice. They also investigate the critical behavior of the corresponding T=0 ferromagnet to spin-glass transition, signaled by a breakdown of the magnetization, using finite-size scaling analyses of the magnetization. The triangular Blume-Capel (BC) model have been studied by using Monte-Carlo simulations (MCs) and discussed that the effects of bond randomness on the universality aspects of a two dimensional BC model embedded in the triangular lattice [9]. Žukovič et al. [10] have been examined effects of selective dilution on phase diagrams and ground-state magnetizations of the spin-1/2 Ising antiferromagnetic model by using EFT. They found that the obtained results showed a fairly good agreement with the previous studies by different method. Low-temperature magnetization processes in a stacked triangular Ising antiferromagnet are studied by utilizing MCs, in detail [11] and observed that multiple steps and hysteresis corresponding to formation of different metastable states in increasing and decreasing magnetic fields. Borovský et al. [12] have studied critical and tricritical behaviors of a selectively diluted triangular Ising antiferromagnet in a field by using both EFT and MCs. Žukovič and Bobák [13] have investigated phase transitions in a triangular BC antiferromagnet and found two kinds of phases. Žukovič and Bobák [14] also studied the critical behaviors of a geometrically frustrated spin-1 Ising antiferromagnet on a triangular lattice in the presence of a single-ion anisotropy by employing MCs. Moreover, by using position-space renormalization group (PSRG) [15] and transfer matrix [16] methods, a frustrated antiferromagnetic spin-1 BC model on a triangular lattice has been examined and found to display a finite-temperature antiferromagnetic (AF) LRO of the type (1, -1, 0) within a certain range of the single-ion anisotropy strength, accompanied with a multicritical behavior.

---
[*] Corresponding author. Tel. +90 352 2076666; Fax: +90 352 4374901.
E-mail addresses: mehmetertas@erciyes.edu.tr (Mehmet Ertaş)

In spite of these studies, to the best of our knowledge, magnetization, susceptibility, phase diagram and hysteresis behaviors of the BC model in a site diluted triangular lattice by means of the EFT with correlations have not been investigated, in detail. Therefore, in this paper, we have studied the influences of the exchange interaction (J), crystal field (D), concentration (p) and temperature (T) on the magnetic properties of spin-1 BC model in a triangular lattice. We should also mention that the ferromagnets have wide applications such as a "soft" ferromagnet in transformer core and a "hard" permanent magnets in hard disk, magnetic tape and motors depending on the extent of the hysteresis loop [16] and [17]. The further development of materials with hysteresis needs a deep understanding of their microscopic interactions and how these interactions influence their hysteresis phenomena [18].

The paper is arranged as follows. In Section 2, we give the model and present the formalism of the model in the EFT. The detailed numerical results and discussions are presented in Section 3. Finally Section 4 is devoted to a summary and a brief conclusion.

## 2. MODEL AND FORMALISM

The Hamiltonian of the selectively diluted triangular lattice includes nearest neighbors interactions and the crystal field is given as:

$$H = -J\sum_{\langle ij \rangle} S_i S_j - D\sum_i S_i^2 - h\sum_i S_i \quad (1)$$

where $S_i$ is the Ising spin and it takes $S_i = 0, \pm 1$ values, D is the crystal field or single-ion anisotropy, h is the external magnetic field, J is the exchange interaction parameter and $\langle i,j \rangle$ denotes the summation over all nearest neighbor pairs.

Within the framework of the EFT with correlations, one can easily find the sublattice magnetizations, and the quadrupolar moment terms as coupled equations for the diluted triangular lattice as follows:

$$m_A = \left[ p\left(1 + m_B \sinh(J\nabla) + m_B^2(\cosh(J\nabla)-1)\right) + 1-p \right]^3 \left[ p\left(1 + m_C \sinh(J\nabla) + m_C^2(\cosh(J\nabla)-1)\right) + 1-p \right]^3 F(x)\Big|_{x=0}, \quad (2a)$$

$$m_B = \left[ p\left(1 + m_A \sinh(J\nabla) + m_A^2(\cosh(J\nabla)-1)\right) + 1-p \right]^3 \left[ p\left(1 + m_C \sinh(J\nabla) + m_C^2(\cosh(J\nabla)-1)\right) + 1-p \right]^3 F(x)\Big|_{x=0}, \quad (2b)$$

$$m_C = \left[ p\left(1 + m_A \sinh(J\nabla) + m_A^2(\cosh(J\nabla)-1)\right) + 1-p \right]^3 \left[ p\left(1 + m_B \sinh(J\nabla) + m_B^2(\cosh(J\nabla)-1)\right) + 1-p \right]^3 F(x)\Big|_{x=0}, \quad (2c)$$

$$q_A = \left[ p\left(1 + m_B \sinh(J\nabla) + m_B^2(\cosh(J\nabla)-1)\right) + 1-p \right]^3 \left[ p\left(1 + m_C \sinh(J\nabla) + m_C^2(\cosh(J\nabla)-1)\right) + 1-p \right]^3 G(x)\Big|_{x=0}, \quad (3a)$$

$$q_B = \left[ p\left(1 + m_A \sinh(J\nabla) + m_A^2(\cosh(J\nabla)-1)\right) + 1-p \right]^3 \left[ p\left(1 + m_C \sinh(J\nabla) + m_C^2(\cosh(J\nabla)-1)\right) + 1-p \right]^3 G(x)\Big|_{x=0}, \quad (3b)$$

$$q_C = \left[ p\left(1 + m_A \sinh(J\nabla) + m_A^2(\cosh(J\nabla)-1)\right) + 1-p \right]^3 \left[ p\left(1 + m_B \sinh(J\nabla) + m_B^2(\cosh(J\nabla)-1)\right) + 1-p \right]^3 G(x)\Big|_{x=0}, \quad (3c)$$

where $\nabla = \partial/\partial x$ is the differential operator. p represents the mean concentration of magnetic atoms. When p probability equal to 1 then the site i is occupied by a magnetic atom and 0 with probability $1-p$ otherwise. The functions $F(x)$ and $G(x)$ are defined as

$$F(x) = \frac{2\sinh[\beta(x+h)]}{\exp(-\beta D) + 2\cosh[\beta(x+h)]} \quad (4a)$$

$$G(x) = \frac{2\cosh[\beta(x+h)]}{\exp(-\beta D) + 2\cosh[\beta(x+h)]} \quad (4b)$$

Here, $\beta = 1/k_B T$, T is the absolute temperature and $k_B$ is the Boltzmann constant. By using the definitions of the order parameters in Eqs. (2a)-(2c), the total $(m_T)$ magnetizations of per site can be defined as $m_T = 1/3(m_A + m_B + m_C)$.

In order to obtained susceptibilities of the system, we differentiated magnetizations respect to h as following equation:

$$\chi_\alpha = \lim_{\to 0}\left(\frac{\partial m_\alpha}{\partial h}\right) \quad (5)$$

where, $\alpha$ = A, B and C. By using of Eqs. (2) and (5), we can easily obtain the $\chi_A$, $\chi_B$ and $\chi_S$ susceptibilities as follow:

$$\chi_A = a_1 \chi_B + a_2 \chi_C + a_3 \frac{\partial F(x)}{\partial h}, \quad (6a)$$

$$\chi_B = b_1 \chi_A + b_2 \chi_C + b_3 \frac{\partial F(x)}{\partial h}. \quad (6b)$$

$$\chi_C = c_1 \chi_A + c_2 \chi_B + c_3 \frac{\partial F(x)}{\partial h}. \quad (6c)$$

Here, $a_i$, $b_i$ and $c_i$ (i=1, 2 and 3) coefficients have complicated and long expressions, hence they will not give. The total susceptibility of per site can be obtain via $\chi_T = 1/3(\chi_A + \chi_B + \chi_C)$.

On the other hand, in order to obtain the second order phase transition temperatures as well as the phase diagram, we must expand the right-hand sides of the (2a)-(2c) coupled equations. They are obtained as follows:

$$Am = \begin{pmatrix} 0 & k_1 & k_2 \\ m_1 & 0 & m_2 \\ n_1 & n_2 & 0 \end{pmatrix} \begin{pmatrix} m_A \\ m_B \\ m_C \end{pmatrix} = 0, \quad (7)$$

Here, the coefficients $k_i$, $m_i$ and $n_i$ in each matrix take complicated forms, so that they will not give. These coefficients can be easily obtained from the coupled equations via differential operator technique. The second order phase transition temperatures of each system can be determined from $det(A) = 0$. Moreover, for the obtaining the first-order phase transition temperatures, we have to solve the Eq. (2) numerically. By this way, we can obtain the tricritical point (TCP) by using the existence of the first- and second-order phase transition points.

Solving these equations, we can get the numerical results of the spin-1 diluted triangular lattice. We will perform these results in the next Section.

## 3. NUMERICAL RESULTS AND DISCUSSIONS

In this section our attention is focused on the study of the magnetic properties, the phase diagrams and hysteresis behavior of a spin-1 Triangular Ising system (TIS) with a crystal field interaction.

### 3.1. Magnetic Properties of Magnetizations and Susceptibilities

The thermal behavior of the total magnetizations ($M_T$) and susceptibilities ($\chi_T$) of the spin-1 TIS are plotted in Fig. 2. In Fig. 2(a)-2(c), we investigated behavior of first and second order phase transition points, as well as tricritical point, for different values of the crystal field, the bilinear interaction parameter and the concentration. Thus, this study leads us to characterize of transitions as well as to obtain the transition points. Fig. 2(a) is obtained for fixed values $J = 1.0$ and $p = 1.0$, and the selected values of D, i.e., -4.0, -3.0, -2.8, -2.0 and -1.0. In this figure for $D = -4.0$, $-3.0$ and $-2.8$ values, the total magnetizations go to zero discontinuously as the temperature increases; hence, a first-order phase transition occurs. One other hand, for $D = -2.0$ and $-1.0$ values the total magnetizations decrease continuously with the increasing of the values of temperature below the critical temperature and they become zero; hence, a second-order phase transition occurs. Moreover, Figs. 2(b) and 2(c) for low values of J and p, namely $J = 0.25$ and $0.35$ values and $p = 0.75$, $0.80$, and $0.85$ values, the system illustrates first order phase transitions; the system exhibits a second-order phase transition for high values of J and p. In the system, the critical values are obtained as $D_C = -2.85$, $J_C = 0.38$, and $p_C = 0.88$, for crystal field, bilinear interaction parameter and the concentration. The susceptibilities diverge as the temperature approaches the critical temperature in Fig. 2. When the D, J, and p take higher values, the susceptibility divergence at the critical temperature shifts to higher temperatures. However, when the D, J, and p take lower values, they become finite and display a jump singularity behavior at the first-order phase transition.

### 3.2. Phase Diagrams

In this subsection, we will show some typical results for the TIS with a crystal field. We have obtained the phase diagrams two different planes, namely (D, T) and (p, T) for TIS.

#### 3.2.1. Phase Diagrams in (D, T) plane

At first, we present the phase diagrams of the model in the (D, T) plane, illustrated in Fig. 3. In these phase diagrams, the solid and dashed lines represent the second- and first-order phase transition lines, respectively, and the tricritical points are denoted by filled circles. It is clear that the second- and first-order phase transition lines separate the ordered phases, namely (1, 1, 1), (-1, -1, 1) and (-1, 0, 1) from the paramagnetic (P) phase. From these phase diagram the following phenomena have been observed. (1) Each one of the phase diagrams exhibits only one tricritical point where the second-order phase transition turns to a first-order one. (2) In Fig. 3(a), the reentrant behavior exists in the TIS, i.e., the system will be disordered (paramagnetic) phase at very low temperatures and as the temperature increases the system antiferromagnetic (1, 1, 1) phase at a critical temperature Tc1 and finally the paramagnetic phase at a higher critical temperature Tc2. (3) In Fig. 3(b), in the system observed the special critical points, namely critical end point (E) and tetracritical point (M).

#### 3.2.2. Phase Diagrams in (p, T) plane

In Fig. 4, the phase diagram of spin 1 TIS is obtained to examine the influence of the interfacial coupling, namely J. Fig. 4 shows the variations of T as a function of p, when the parameter J is fixed at J = 1.0 and the value of D is changed (D = 0.0, -1.0, -1.5 and -2.0). In Fig. 4, the phase transition region is divided into two phases, namely P and (1, 1, 1). From Fig. 4, we can see that the phase transitions are the second-order and first order phase transitions for high and low values of temperature, respectively. When the cristal field take higher values, the tricritical points are obtained in the lower temperature.

### 3.3. Hysteresis behaviors

Our investigations in this subsection are to examine the effects of the crystal field, concentration, bilinear interaction, and temperatures on the hysteresis behaviors of a spin-1 TIS. Fig. 5(a) is obtained for selected four typical concentration values, namely p = 0.3, 0.6,

0.8, 1.0 in the case of D = -2.0, T = 0.5 and J = 1.0 fixed values to investigate the temperature dependence of the hysteresis and susceptibility behaviors of the spin-1 TIS. With the increase of crystal field parameter, the double hysteresis loop turn to single hysteresis loop and the single hysteresis loop becomes wider. In Fig. 5(b), the similar hysteresis loop behaviors with Fig. 5(a) are observed for the increasing values of D. The total susceptibility peaks confirm the above calculations. In Fig. 5(c), the hysteresis loops area increase as the bilinear parameter increases. This fact is also understand from the susceptibility peak turn to reaches two values in both directions of external magnetic field. Fig. 5(d), the hysteresis loops area decrease as the temperature increases. This fact is also understand from the susceptibility peak turn to reaches a single value in both directions of external magnetic field. The physical means of this fact is that while at low temperatures the system becomes hard magnet, with the increase of the temperature the hard magnet turn to soft magnet. These results are consistent with some experimental results [19-22].

## 4. SUMMARY AND CONCLUSION

In this study, we studied some magnetic behaviors of Blume-Capel (BC) model in a site diluted triangular lattice by means of the effective-field theory (EFT) with correlations. We investigated the magnetic properties of spin-1 BC model in a triangular lattice in detail. We also obtained the phase diagrams of the system in two different planes. The tricritical point as well as tetracritical and critical end special points are found as depending on the physical parameters of the system. Moreover, when the hysteresis behaviors of the system are examined, the single and double hysteresis loop are observed for various values of the physical parameters. We show that the hysteresis loops have different coercive field points in which the susceptibility make peak at these points.

## ACKNOWLEDGEMENTS

This work was supported by the Scientific and Technological Research Council of Turkey (TUBITAK) (grant no: 114F008).

**FIGURE CAPTIONS**

**FIG. 1.** (Color online) Schematic representation of triangular Ising system. The red, blue, and green spheres indicate magnetic atoms at the $m_A$, $m_B$, $m_C$, respectively.

**FIG. 2.** (Color online) Thermal variations of the magnetizations and susceptibilities with the various values of p, J and D.

**(a)** p = 1.0, J = 1.0 and D = -4.0, -3.0, -2.8, -2.0, -1.0.
**(b)** p = 1.0, D = 1.0 and J= 0.25, 0.35, 0.5, 0.75, 1.0.
**(c)** J=1.0, D = 1.0 and p = 0.75, 0.8, 0.85, 0.89, 1.0.

**FIG. 3.** The phase diagrams in (T-D) plane of triangular Ising system. Dashed and solid lines represent the first- and second-order phase transitions, respectively.

**(a)** J = 1.0 and p = 1.0.
**(b)** J = -1.0 and p = 1.0.
**(c)** J = -1.0 and p = 0.5.

**FIG. 4.** The phase diagrams in (T-p) plane of triangular Ising system. Dashed and solid lines represent the first- and second-order phase transitions, respectively. For J = 1.0 and D = 0.0, -1.0, -1.5, -2.0.

**FIG. 5.** (Color online) Hysteresis behaviors of triangular Ising system.

    **(a)** $D = -2.0$, $J = 0.5$, $T = 1.0$ and $p = 0.3, 0.6, 0.8, 1.0$.

    **(b)** $J = 1.0$, $T = 0.5$, $p = 0.5$ and $D = -3.0, -2.0, -1.0, 1.0$.

    **(c)** $D = 0.0$, $T = 0.5$, $p = 0.5$ and $J = 0.1, 0.5, 0.7, 1.0$.

    **(d)** $D = 0.0$, $J = 0.5$, $p = 0.5$ and $T = 0.1, 0.5, 0.7, 1.1$.

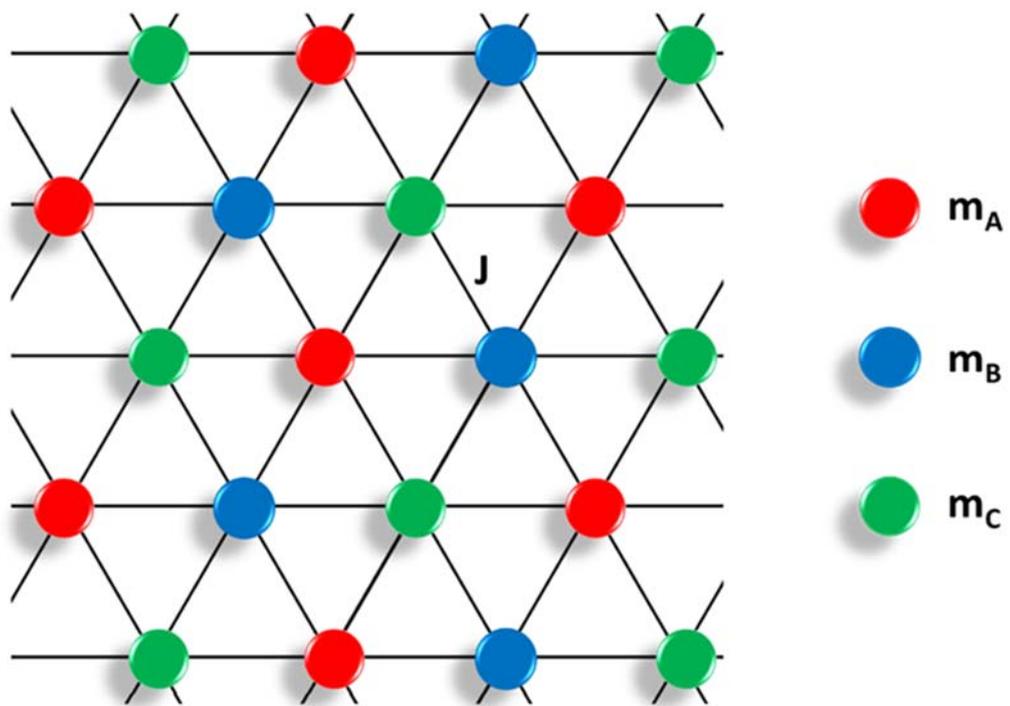

**Fig. 1**

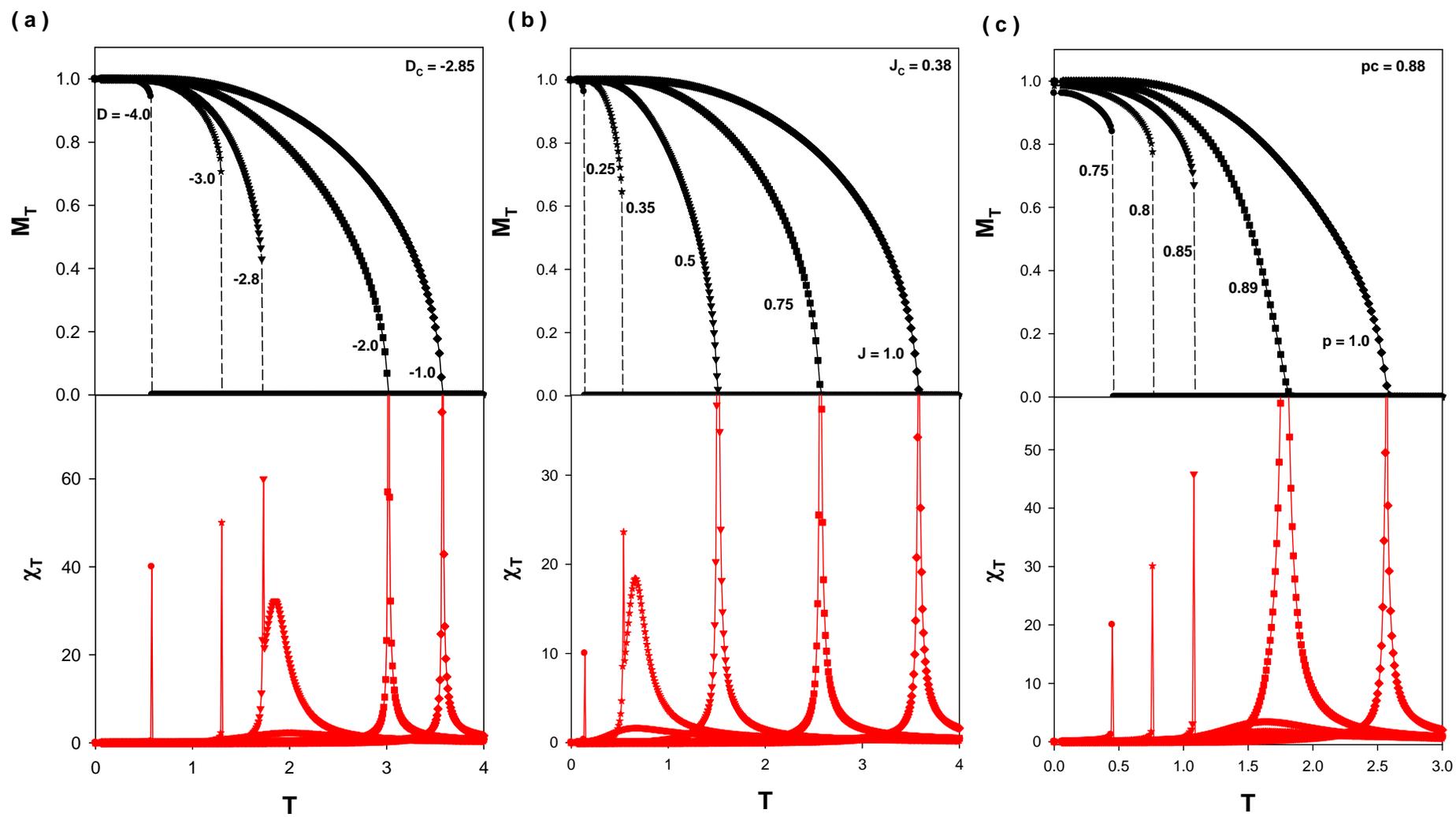

FIG. 2

Figure 3

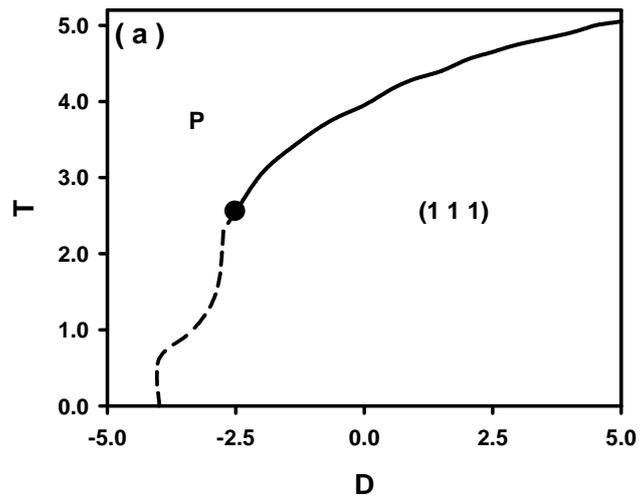
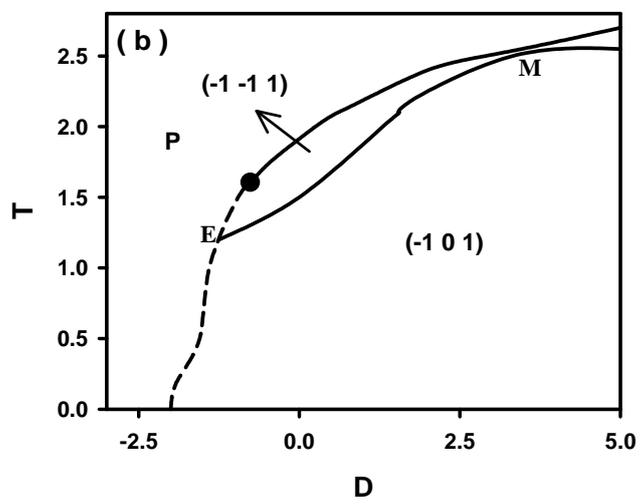
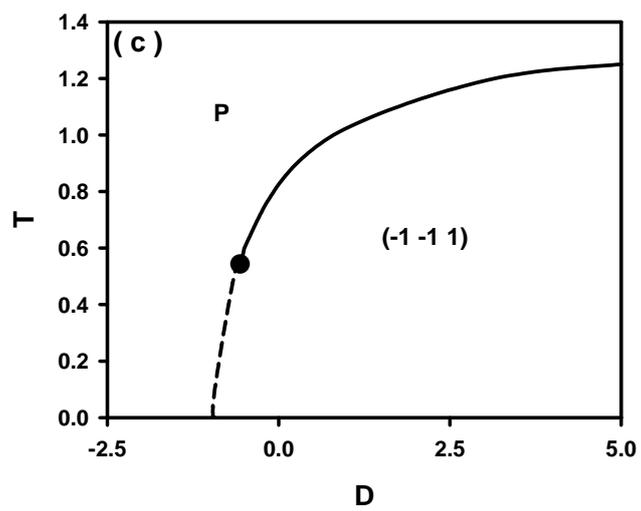

FIG. 3

**Figure 4**

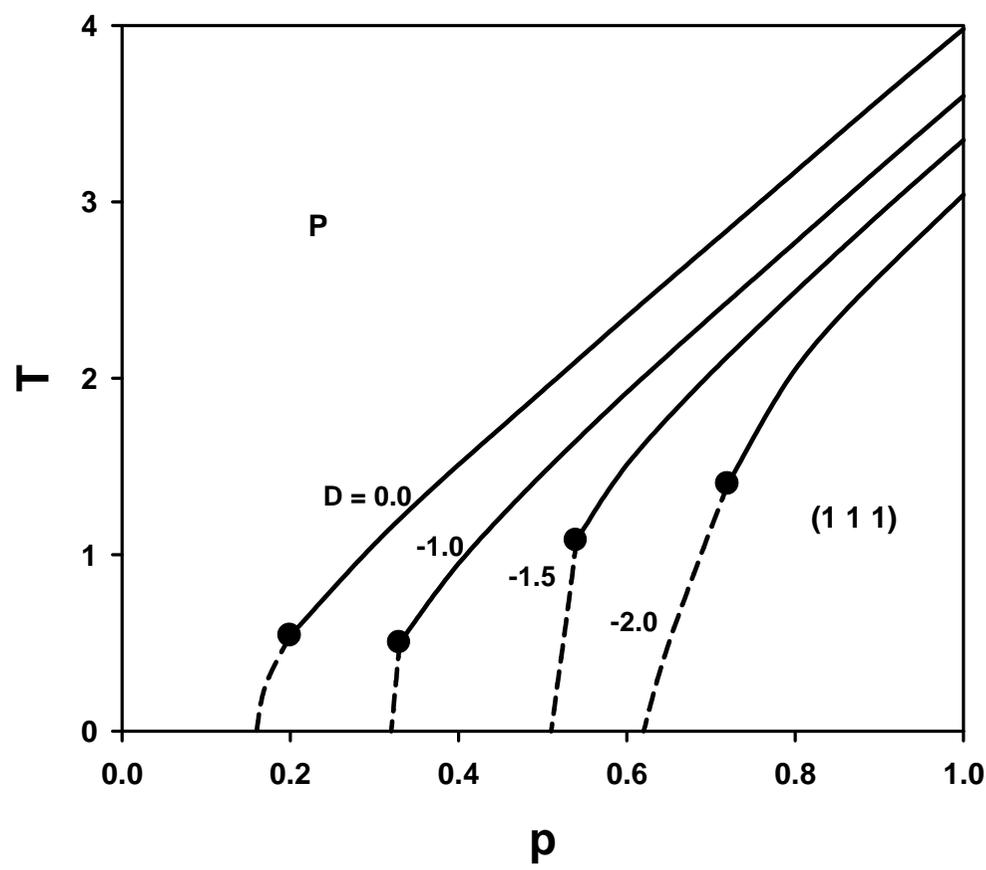

**FIG. 4**

**Figure 5a**

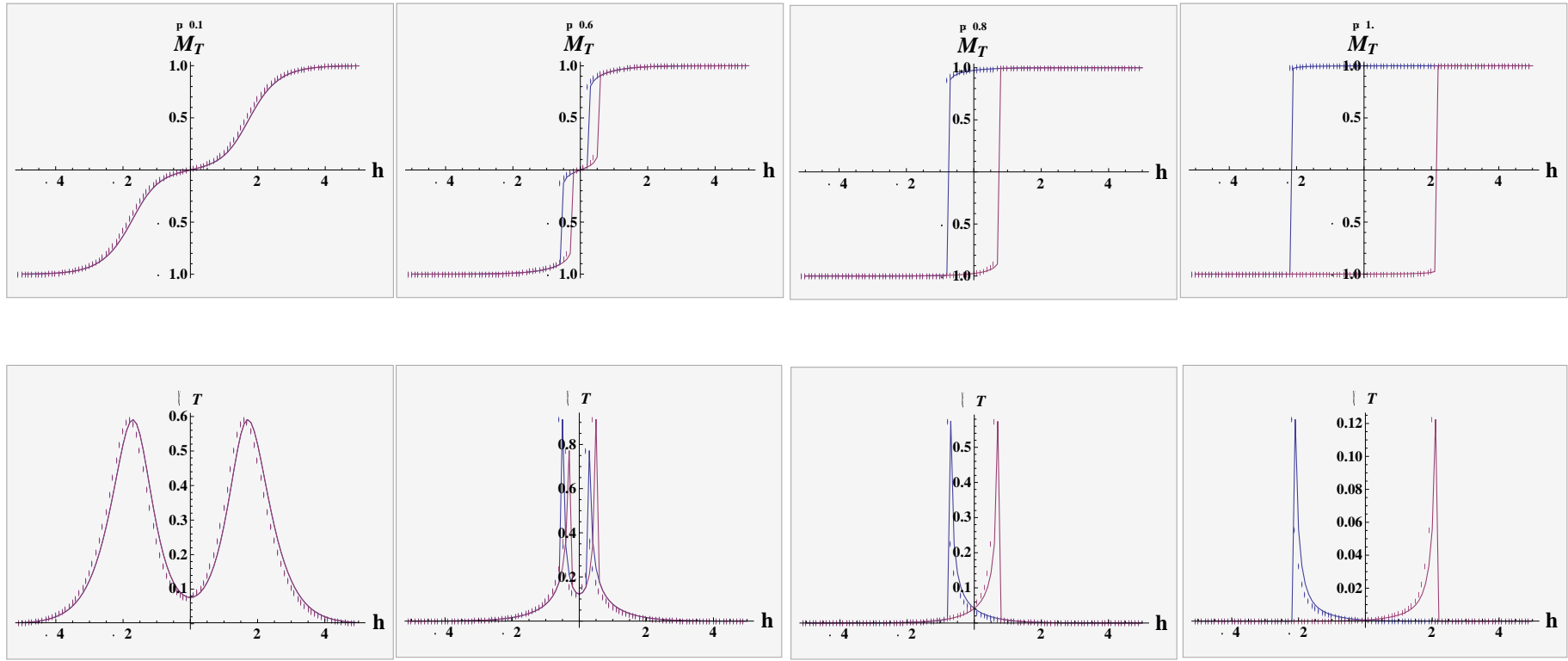

**FIG. 5(a)**

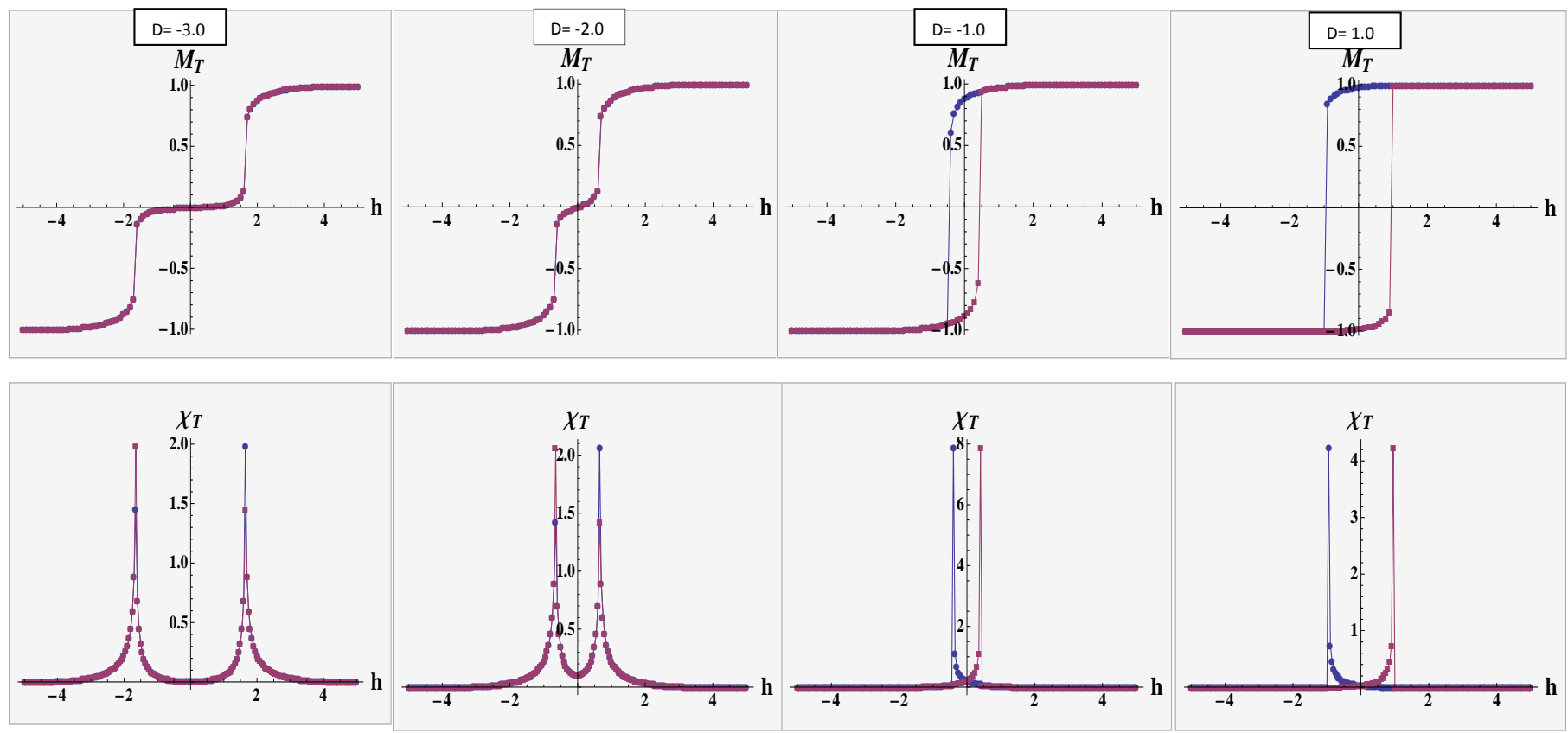

FIG. 5(b)

">Figure 5c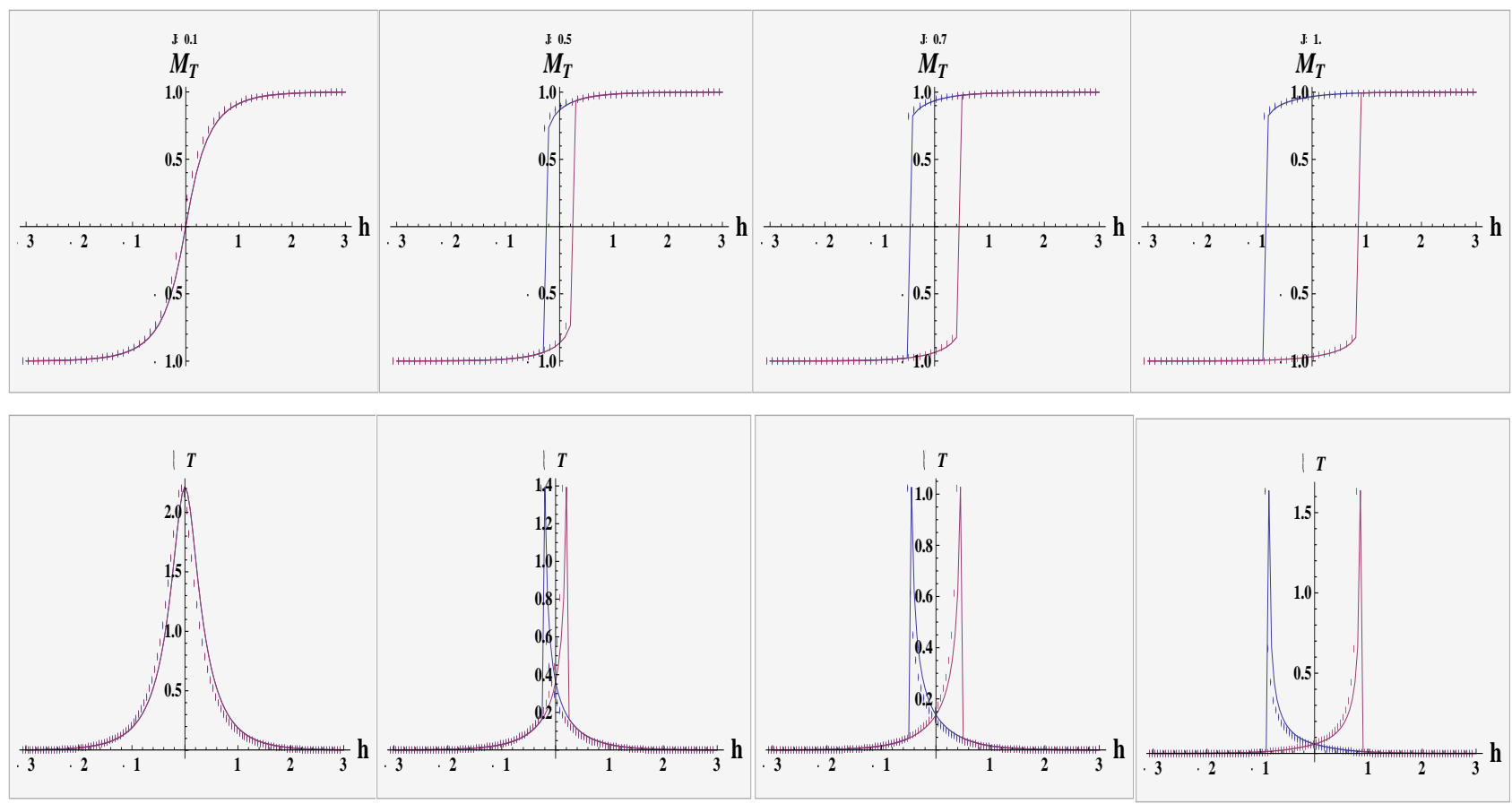

**FIG. 5(c)**

**Figure 5d**

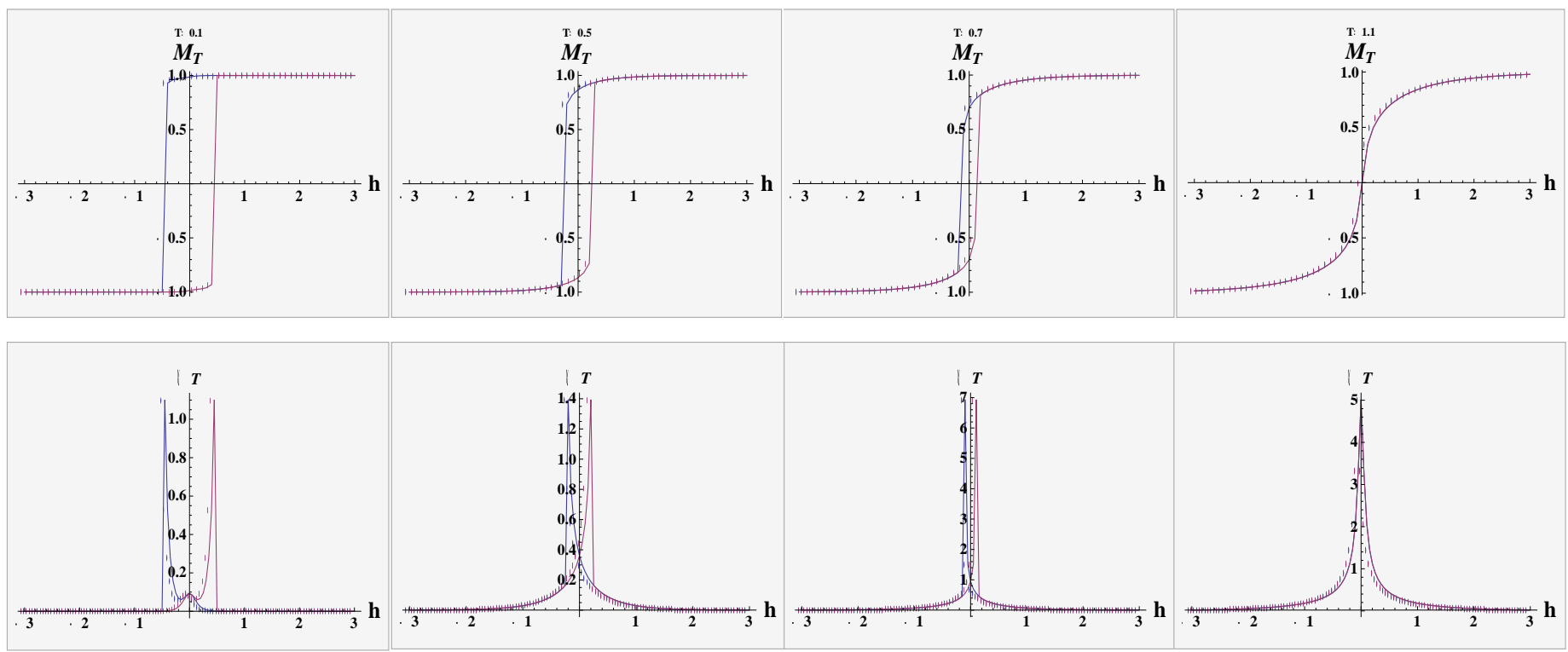

**FIG. 5(d)**